# Fullerene-rare gas mixed plasmas in an electron cyclotron resonance ion source[a]

T. Asaji,[1,b] T. Ohba,[1] T. Uchida,[2] H. Minezaki,[3] S. Ishihara,[3] R. Racz,[4] M. Muramatsu,[5] S. Biri,[4] A. Kitagawa,[5] Y. Kato,[6] and Y. Yoshida[2]

[1]Oshima National College of Maritime Technology, 1091-1 Komatsu, Suo-oshima, Oshima, Yamaguchi 742-2193, Japan

[2]Bio-Nano Electronics Research Centre, Toyo University, 2100 Kujirai, Kawagoe, Saitama 350-8585, Japan

[3]Graduate School of Engineering, Toyo University, 2100 Kujirai, Kawagoe, Saitama 350-8585, Japan

[4]Institute of Nuclear Research (ATOMKI), H-4026 Debrecen, Bem Tér 18/c, Hungary

[5]National Institute of Radiological Sciences (NIRS), 4-9-1 Anagawa, Inage-ku, Chiba 263-8555, Japan

[6]Graduate School of Engineering, Osaka University, 2-1 Yamada-oka, Suita, Osaka 565-0871, Japan

A synthesis technology of endohedral fullerenes such as Fe@$C_{60}$ has developed with an electron cyclotron resonance (ECR) ion source. The production of N@$C_{60}$ was reported. However, the yield was quite low, since most fullerene molecules were broken in the ECR plasma. We have adopted gas-mixing techniques in order to cool the plasma and then reduce fullerene dissociation. Mass spectra of ion beams extracted from fullerene-He, Ar or Xe mixed plasmas were observed with a Faraday cup. From the results, the He gas mixing technique is effective against fullerene destruction.

## I. INTRODUCTION

We have developed a synthesis technology of endohedral fullerenes such as Fe-$C_{60}$ (Fe@$C_{60}$) with an electron cyclotron resonance ion source (ECRIS). We succeeded in the production of endohedral N@$C_{60}$ and Fe@$C_{60}$ by an ion irradiation method.[1,2] The yields were quite low because only the surfaces of $C_{60}$ thin films were irradiated with ion beams. We have attempted gas-phase synthesis in the ECR plasma as an alternative. One of the authors has been succeeded in synthe-sizing N@$C_{60}$ in an ECR plasma.[3] However, most fullerene molecules were broken in the plasma. High-temperature electrons probably damaged the fullerenes and endohedral ones.

ECR plasmas under low-pressure conditions produce easily high-temperature electrons compared with dissociation and ionization energies of fullerenes, i.e., 7−12 eV.[4] For example, we reported that the maximum electron temperature at a microwave power of 40 W were approximately 9 eV in the 2.45 GHz ECRIS.[5] On the other hand, microwave power required for iron plasma generation was much higher than that required for fullerene plasma generation. Since the conditions for both plasmas differ considerably, it is difficult to generate the dense mixed plasma at present. Therefore, we need to develop a technique not to damage fullerenes at higher microwave power.

In our previous work, we studied an effect of pulse-modulated microwaves on the dissociation of fullerene molecules.[6] The pulse modulation technique can decrease electron temperatures while maintaining high plasma density. Contrary to our expectation, broken fullerenes in the plasma were increased by the technique. The increase was probably caused by an increase in the peaks of the microwave power although the average power remained constant.

To reduce the fullerene dissociation, we have adopted gas-mixing techniques. The techniques have been studied in ECRIS field for 30 years.[7,8] The effects described in Ref. 8 are as follows: (i) a dilution effect lowering the mean ion charge, (ii) ion cooling resulting from the mass effect in ion ion collisions, (iii) an increase of the electron density because of the better ionization efficiency of the added gas, and

(iv) an increase of the plasma stability. The techniques have been developed to improve the production of highly charged ions. Our group[9] and Maunoury et al.[10] also reported that the compositions in fullerene plasmas were changed by gas mixing. Therefore, we thought of adopting the gas mixing to reduce the fullerene dissociation because microwave power is mainly used for not the dissociation but rather mixing gas ionization. Additionally, improving confinement of fullerene ions is expected by the cooling effect.

In this paper, we investigated fullerene and rare-gas mixed plasmas. The mass spectra of the extracted ion beams are shown and we discuss the effect of the gas mixing on fullerene dissociation.

## II. EXPERIMENT

Schematic diagrams of the Bio-Nano ECRIS are shown in Fig. 1. The details have been described in Ref. 6. Fullerene powder used in this experiment was four nines $C_{60}$ (Frontier Carbon Co., nanom purple Su). The fullerene vapor was introduced with a filament oven. The temperature was about 600 K. Mixing gases were helium, argon, and xenon. The flow rates were set at 0.28 or 0.56 SCCM (SCCM denotes cubic centimeters per minute at standard temperature and pressure). The output frequency of a microwave source was

---

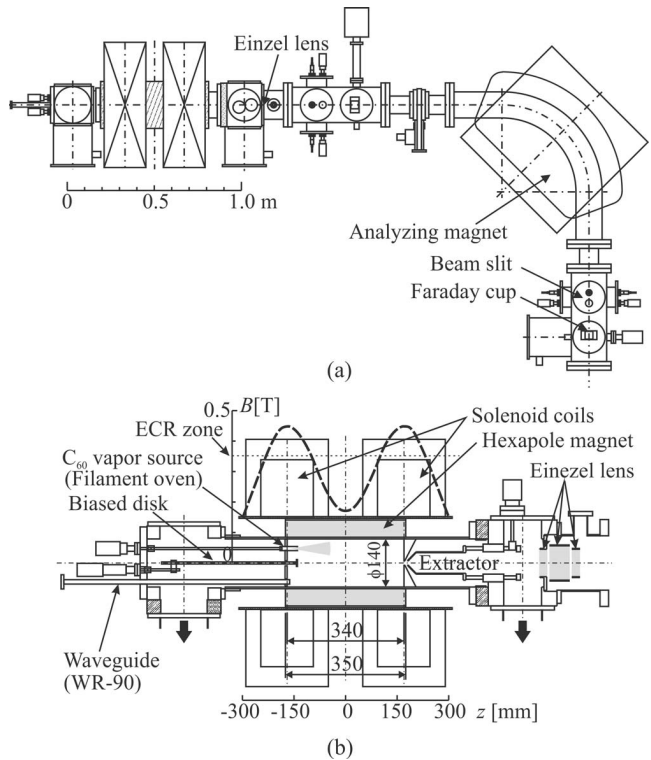

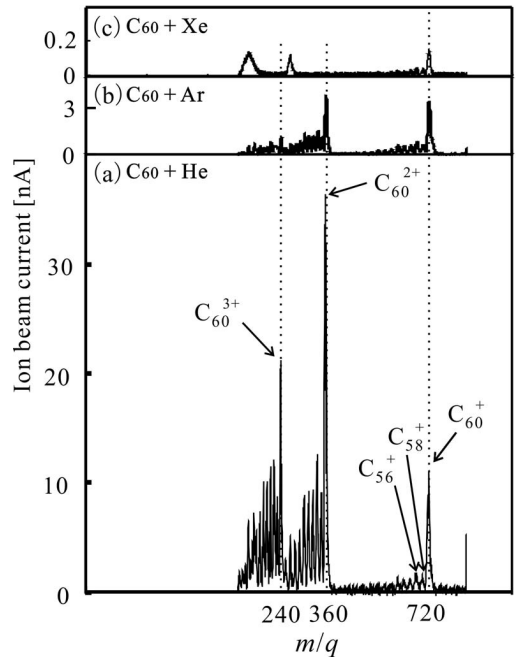

FIG. 1. Schematic diagrams of the Bio-Nano ECRIS. (a) Bio-Nano ECRIS apparatus. (b) ECR ion source.

FIG. 3. Mass spectra of ion beams extracted from fullerene-(a) helium, (b) argon or (c) xenon mixed plasmas. The flow rate of each rare gas is 0.28 SCCM. The microwave power and frequency were 4 W and 9.75 GHz, respectively. The pressures were (a) $5.9 \times 10^{-5}$, (b) $1.1 \times 10^{-4}$, and (c) $1.1 \times 10^{-3}$ Pa.

set at 9.75 GHz. The power was varied between 2 and 8 W. Each current of mirror coils was 500 A. Mass spectra were recorded with a Faraday cup after a bending magnet.

## III. RESULTS AND DISCUSSION

As a fundamental study, we reported the responses of ion saturation currents to pulse modulated microwaves for fullerene plasma.[6] Additionally, the pulse responses for He, Ar, and Xe plasmas have been measured with a biased disk set to $-58$ V (Fig. 2). The time shift between the pulse response for Ar and the pulse signal is caused by microwave reflection on the leading edge of the pulse. A rise time for Xe is fast compared with those for He and Ar. The ion saturation current for Xe is 2–4 times higher than those for the others. These differences can be attributed to the ionization energies. Those of He, Ar, and Xe are 24.6, 15.8, and 12.1 eV, respectively. Therefore, Xe as a mixing gas can raise electron density under low microwave power conditions. As He is more difficult to be ionized due to a high first ionization energy, that is not suitable for a mixing gas to increase the density of fullerene plasmas. It is the opposite for Xe.

Figure 3 shows mass spectra of ion beams extracted from fullerene (a) He, (b) Ar, and (c) Xe mixed plasmas. The flow rate of the added gases was 0.28 SCCM. The pressures were (a) $5.9 \times 10^{-5}$, (b) $1.1 \times 10^{-4}$, and (c) $1.1 \times 10^{-3}$ Pa. The microwave power was 4 W in CW mode. The fullerene beam currents for He are much higher than those for Ar and Xe. The result indicates that He gas cannot break easily the fullerenes. For Xe mixing, undissociated fullerene ions, e.g., $C_{60}^+$ and $C_{60}^{2+}$, were hardly observed. In the case of 2 W (not shown), these ion currents slightly increased. For Ar mixing, the peaks of the fullerene ions can be seen. The ion beam currents are $3-10$ times lower than those for He mixing. The degree of the fullerene destruction can be explained by differences of momentum. The atomic weights of He, Ar, and Xe are approximately 4, 40, and 131 g/mol, respectively. Assuming that these ion temperatures are equal, the momentum affecting collisions is proportional to the square root of the weight. Thus, the momentum for Ar and Xe is 3.2 and 5.7 times higher than that for He.

Hence, He gas can hardly damage fullerene molecules and carry off the energies of fullerenes as shown by the cooling effect. It is clear that heavy gases cannot play the role. On the other hand, dissociation ratios such as $C_{58}^+/C_{60}^+$ rarely vary with the kind of added gases.

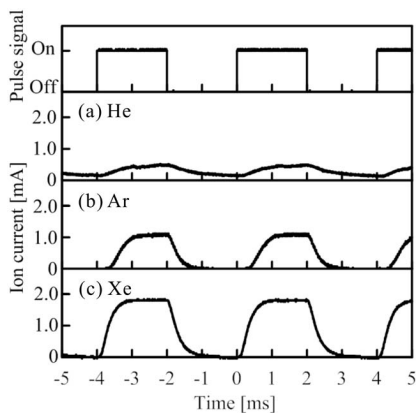

FIG. 2. The responses of ion saturation currents measured with the biased disk to the signal of pulse modulation for (a) helium, (b) argon, and (c) xenon plasmas.

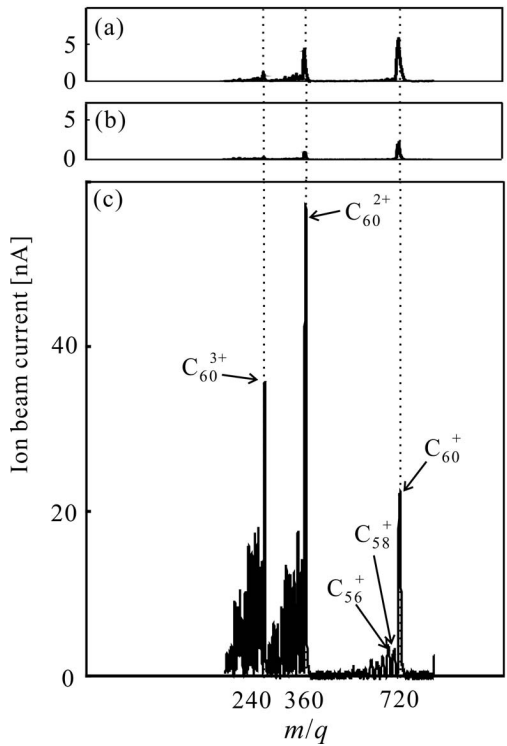

FIG. 4. Mass spectra of ion beams extracted from fullerene-helium mixed plasmas. The microwave power, the helium flow rates and the pressures were (a) 2 W, 0.28 SCCM, and $5.7 \times 10^{-5}$ Pa, (b) 4 W, 0.56 SCCM, and $2.6 \times 10^{-4}$ Pa, and (c) 8 W, 0.28 SCCM, and $5.9 \times 10^{-5}$ Pa, respectively.

Mass spectra of ion beams extracted from $C_{60}$-He plasmas investigated under several conditions [Figs. 4(a)–4(c)]. For a microwave power of 2 W (a), i.e., half the power of the above condition [Fig. 3(a)], $C_{60}^+$ beam current decreased by half and $C_{60}^{2+}$ beam current dropped steeply. The result can be explained by a decrease in plasma density with microwave power. For He flow rate of 0.56 SCCM (b), i.e., the double flow rate, $C_{60}^+$ and $C_{60}^{2+}$ beam currents were hardly observed. The chamber pressures for 0.28 and 0.56 SCCM were $5.7$–$5.9 \times 10^{-5}$ and $2.6 \times 10^{-4}$ Pa, respectively. This tendency probably results from a decrease in the electron temperature with increasing collisions in the plasma. For a microwave power of 8 W (c), i.e., the double power, $C_{60}^+$, $C_{60}^{2+}$, and $C_{60}^{3+}$ beam currents increased with increasing the power.

The result shows no indication of enhancing fullerene dissociation. In these conditions, the beam currents of $C_{60}^+$, $C_{60}^{2+}$, and $C_{60}^{3+}$ are approximately proportional to the power. We have not yet conducted the experiments at higher microwave power. However, we show the possibility of the synthetic experiment at a high power compared to our previous works.

It is clear that the He gas mixing technique is effective against fullerene destruction. However, there are not many candidate gases, because atoms to bind easily with carbon are not suitable as a mixing gas. Therefore, we will attempt to synthesize Fe@$C_{60}$ in Fe-fullerene plasma added He gas.

## ACKNOWLEDGMENTS

This work was supported by JSPS KAKENHI Grant Nos. 24810029 and 24710095. The participation of two of the authors (S.B., R.R.) in this work was partly supported by the TAMOP 4.2.2.A-11/1/KONV-2012-0036 project, which is co-financed by the European Union and European Social Fund. Part of this study has been supported by a Grant for the Strategic Development of Advanced Science and Technology S1101017 organised by the Ministry of Education, Culture, Sports, Science and Technology (MEXT) since April 2011.